\documentclass{PoS}
\usepackage{wrapfig}

\newcommand{\be}{\begin{equation}}
\newcommand{\ee}{\end{equation}}
\newcommand{\bea}{\begin{eqnarray}}
\newcommand{\eea}{\end{eqnarray}}
\newcommand{\msbar}{{\overline{{\rm MS}}}}

\title{$K \to \pi$ matrix elements of the chromagnetic operator on the lattice}

\ShortTitle{$K \to \pi$ matrix elements of the chromagnetic operator on the lattice}

\author{M. Constantinou$^a$, M. Costa$^a$, R. Frezzotti$^b$, \speaker{V. Lubicz}\,$^{c,d}$, G. Martinelli$^e$, D. Meloni$^{c,d}$, H.~Panagopoulos$^a$, S. Simula$^d$ \\ \\
\llap{$^a$} Department of Physics, University of Cyprus, CY-1678 Nicosia, Cyprus\\
\llap{$^b$} Dipartimento di Fisica, Universit\`a di Roma ``Tor Vergata'' and INFN, Sezione ``Tor Vergata'', I-00133 Rome, Italy\\
\llap{$^c$} Dipartimento di Matematica e Fisica, Universit\`a Roma Tre, I-00146 Rome, Italy\\
\llap{$^d$} INFN, Sezione di Roma Tre, I-00146 Rome, Italy\\
\llap{$^e$} SISSA, I-34136 Trieste, Italy

\bigskip
E-mail: \email{constantinou.martha@ucy.ac.cy},
        \email{kosta.marios@ucy.ac.cy}, \email{roberto.frezzotti@roma2.infn.it},
        \email{lubicz@fis.uniroma3.it}, \email{guido.martinelli@sissa.it},
        \email{meloni@fis.uniroma3.it}, \email{haris@ucy.ac.cy}, \email{simula@roma3.infn.it}

\bigskip
{\large\bf ETM Collaboration}}

\abstract{We present preliminary results of the first lattice QCD calculation of the $K\to\pi$ matrix elements of the chromomagnetic operator $O_{CM}=g\, \bar s\, \sigma_{\mu\nu} G_{\mu\nu} d$, which appears in the effective Hamiltonian describing $\Delta S=1$ transitions in and beyond the Standard Model. Having dimension 5, the chromomagnetic operator is characterized by a rich pattern of mixing with operators of equal and lower dimensionality. The multiplicative renormalization factor as well as the mixing coefficients with the operators of equal dimension have been computed at one-loop in perturbation theory. The power divergent coefficients controlling the mixing with operators of lower dimension have been computed non-perturbatively, by imposing suitable subtraction conditions. The numerical simulations have been carried out using the gauge field configurations produced by the European Twisted Mass Collaboration with $N_f = 2+1+1$ dynamical quarks at three values of the lattice spacing. Our preliminary result for the B-parameter of the chromomagnetic operator is $B_{CMO}=0.29(11)$, which can be compared with the estimate $B_{CMO}\sim 1-4$ currently used in phenomenological analyses.}

\FullConference{The 32nd International Symposium on Lattice Field Theory,\\
		23-28 June, 2014\\
		Columbia University New York, NY}

\begin{document}

\section{Introduction}
At low energy with respect to the electroweak scale, the Standard Model (SM) and its possible New Physics (NP) extensions are described by an effective Hamiltonian in which the contribution of operators of dimension $d=4+n$ are suppressed by $n$ powers of the high-energy (i.e. the electroweak or NP) scale. In the flavor changing $\Delta S=1$ sector, the effective Hamiltonian contains four operators of dimension $d=5$, two electromagnetic (EMO) and two chromomagnetic (CMO) operators. Their contribution to the physical amplitudes is thus suppressed by only one power of the high-energy scale. The $\Delta S = 1,\ d=5$ effective Hamiltonian has the form
\be
H^{\Delta S = 1,\ d=5}_{\rm eff} = \sum_{i=\pm} \left(C^i_\gamma \, Q^i_\gamma + C^i_g \, Q^i_g \right) + {\rm h.c.}\ ,
\ee
where $Q_{\gamma,g}^+$ ($Q_{\gamma,g}^-$) are the parity even (odd) EMO and CMO respectively defined as:
\bea
&& Q_\gamma^\pm = {Q_d\,e\over 16 \pi^2} \left(\bar s_{L} \,\sigma^{\mu\nu}\, F_{\mu\nu} \, d_{R} 
\pm \bar s_{R} \,\sigma^{\mu\nu}\, F_{\mu\nu} \, d_{L}  \right) \ , \nonumber \\
&& Q_g^\pm = { g\over 16 \pi^2} \left(\bar s_{L} \,\sigma^{\mu\nu}\, G_{\mu\nu} \, d_{R}   
\pm \bar s_{R} \,\sigma^{\mu\nu}\, G_{\mu\nu} \, d_{L} \right) \ ,
\eea
with $q_{R,L}=\frac{1}{2}(1 \pm \gamma_5)\, q$ (for $q=s,d$).

\begin{figure}[ht!]
\centering
\includegraphics[scale=0.5]{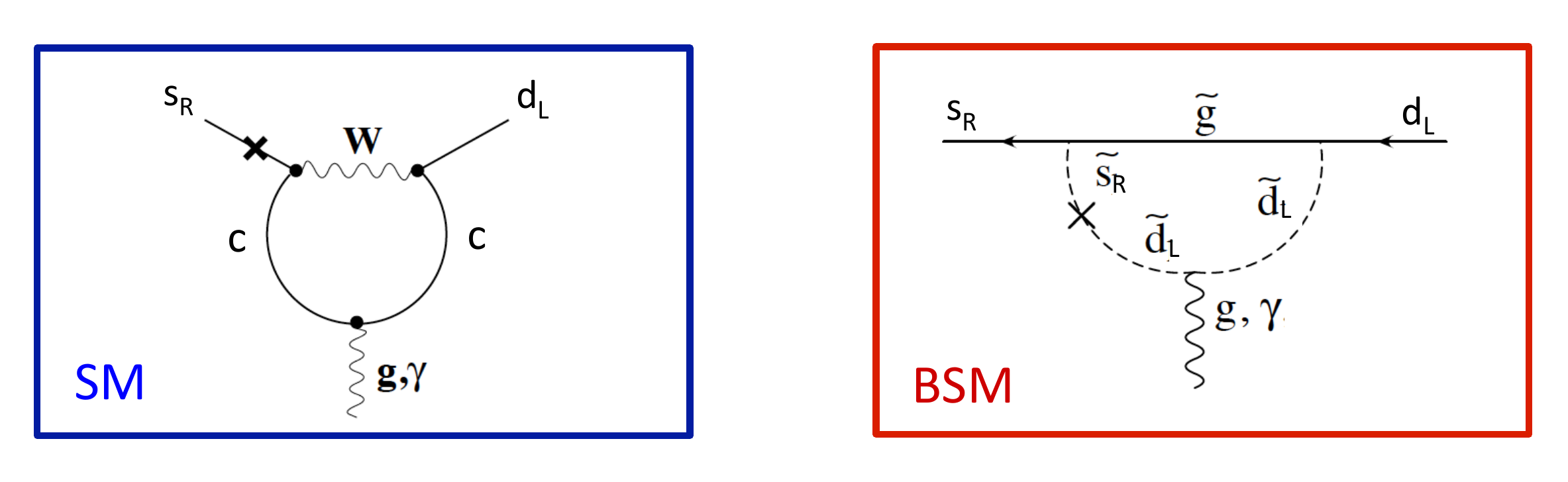}
\caption{\it One-loop Feynman diagrams contributing at low energy to the effective magnetic interactions, in the SM (left) and beyond (right). In the latter case we have shown, for illustrative purposes, the case of SUSY models. The crosses denote a mass insertion.}
\label{fig:fd}
\end{figure}

In Fig.~\ref{fig:fd} we show two examples of Feynman diagrams generating, at low energy, the effective magnetic interactions in the SM and beyond. As NP contribution, we have considered for illustration the case of SUSY models, in which the $\Delta S = 1$ transition occurs through the exchange of virtual squarks and gluinos and it is mediated by the strong interactions.
Note that at least one mass insertion is required in the diagrams, both in the SM and beyond, in order to induce the chirality flip described by the magnetic operators. 

A quick inspection of the diagrams of Fig.~\ref{fig:fd} shows that the Wilson coefficients of the magnetic operators in the SM and NP model are proportional to
\be
C_{\gamma,g}^{SM} \sim \frac{\alpha_w(M_W)}{M_W} \, \frac{m_s}{M_W} \quad , \quad
C_{\gamma,g}^{NP} \sim \frac{\alpha_s(M_{NP})}{M_{NP}} \, \delta_{LR} \ ,
\label{eq:cWilson}
\ee
where $M_{NP}$ represents the typical NP scale, e.g. the gluino mass in the SUSY case, and the factors $m_s/M_W$ and $\delta_{LR}$ are generated in the diagrams by the mass insertion. In the SUSY case, for instance, $\delta_{LR}$ represents the off-diagonal matrix element of the squark mass matrix normalized to the average squark mass. The transition rate is controlled in the SM by the weak coupling $\alpha_w(M_W)$. This is not generally the case for NP models. In the SUSY transition shown in Fig.~\ref{fig:fd}, for example, the process is mediated by the strong interactions. Therefore, the proportionality of $C_{\gamma,g}^{NP}$ in Eq.~(\ref{eq:cWilson}) to the strong coupling constant $\alpha_s(M_{NP})$, rather than to the weak coupling as in the SM, compensates in part for the stronger high-energy scale suppression ($M_{NP} > M_W$) in the NP model. Thus, the magnetic interactions receive potentially large contributions from physics beyond the SM. 

It is also worth noting that the chirality flipping factor $m_s/M_W$, which appears in the Wilson coefficients of the magnetic operators in the SM, is of the same size of $\Lambda_{QCD}/M_W$, which represents the additional suppression factor of the coefficients of dimension-6 operators in the effective Hamiltonian. For this reason, the role of the magnetic operators tends to be marginal in the SM, while it is potentially more relevant for the searches of NP.

The $K \to \pi$ matrix element of the EMO $Q_{\gamma}^+$, which is relevant for instance for the CP violating part of the rare $K_L \to \pi^0 l^+ l^-$ decays~\cite{Buras:1999da}, has been computed on the lattice both in the quenched~\cite{Becirevic:2000zi} and unquenched $N_f=2$~\cite{Baum:2011rm} case. Since the electromagnetic field strength tensor $F_{\mu\nu}$ factorizes out of the hadronic matrix element, the lattice computation only involves the quark bilinear operator $\bar s \,\sigma^{\mu\nu} d$, and it is relatively straightforward. 

Computing the hadronic matrix elements of the CMO $Q_{g}^\pm$ is, instead, by far more challenging. The main difficulty is represented by the complicated renormalization pattern of the operator, which also involves power divergent mixing with operators of lower dimensionality (see Sect. 2). The relevant matrix elements with an initial kaon involve one, two or three pions in the final states, and are of great phenomenological interest for various processes: the long distance contribution to $K^0 - \bar K^0$ mixing~\cite{He:1999bv}, $\Delta I=1/2$, $K\to \pi\pi$ transitions and $\varepsilon^\prime/\varepsilon$~\cite{Buras:1999da}, CP violation in $K\to 3\,\pi$ decays~\cite{D'Ambrosio:1999jh}. These matrix elements are parameterized in terms of suitably defined B-parameters:
\bea
\label{eq:me}
&& \langle \pi^+ | Q_g^+ | K^+ \rangle = \frac{11}{32\pi^2}\, \frac{M_K^2 \, (p_K \cdot p_\pi)}{m_s + m_d} \, B_{CMO}^{K\pi} \ , \nonumber \\
&& \langle \pi^+ \pi^- | Q_g^- | K^0 \rangle = i\, \frac{11}{32\pi^2}\, \frac{M_K^2 \, M_\pi^2}{f_\pi\,(m_s + m_d)} \, B_{CMO}^{K2\pi} \ , \\
&& \langle \pi^+ \pi^+ \pi^- | Q_g^+ | K^+ \rangle = - \frac{11}{16\pi^2}\, \frac{M_K^2 \, M_\pi^2}{f_\pi^2\,(m_s + m_d)} \, B_{CMO}^{K3\pi} \ .\nonumber
\eea

At leading order (LO) in chiral perturbation theory (ChPT), the CMO has a single representation in terms of pseudo-Goldstone boson fields~\cite{Bertolini:1994qk},
\be
\label{eq:chiral}
Q_g^\pm = \frac{11}{256\pi^2}\, \frac{f_\pi^2 \, M_K^2}{m_s + m_d} \, B_{CMO} \, 
\left[ U (D_\mu U^\dagger) (D^\mu U) \pm  (D_\mu U^\dagger) (D^\mu U) U^\dagger \right]_{23} \ ,
\ee
where the low-energy constant $B_{CMO}$ is estimated to be of order 1 in the chiral quark model of Ref.~\cite{Bertolini:1994qk}. Therefore, the three B-parameters of Eq.~(\ref{eq:me}) are related by chiral symmetry, which predicts at LO their equality: $B_{CMO}^{K\pi} = B_{CMO}^{K2\pi} = B_{CMO}^{K3\pi} = B_{CMO}$. In the present study, we take advantage of this prediction and of the chiral representation (\ref{eq:chiral}) of the CMO in order to evaluate the matrix elements of Eq.~(\ref{eq:me}) from the lattice computation of the single, off-shell $\langle \pi | Q_g^+ | K\rangle$ matrix element.
 
\section{Renormalization of the chromomagnetic operator}
The most challenging aspect in the study of the CMO is represented by its renormalization pattern. The specific structure of the mixing depends on the details of the lattice regularization, i.e. on the choice of the lattice action. For this study, we used the gauge configurations produced by the European Twisted Mass Collaboration (ETMC) with $N_f = 2 + 1 + 1$ dynamical quarks, in which gluons are described by the Symanzik improved Iwasaki action~\cite{Iwasaki:1985we} and quarks by the twisted mass action at maximal twist~\cite{Frezzotti:2003ni}. A detailed analysis of the discrete symmetries of this action then shows that the renormalization of the CMO involves the mixing among 13 operators, of equal or lower dimensionality, including also non gauge invariant operators vanishing by the equation of motion. For the on-shell matrix elements the mixing simplifies, and the renormalized parity even CMO is expressed by
\be
\label{eq:reno}
\widehat O_{CM} = Z_{CM} \left[O_{CM} - \left( \frac{c_{13}}{a^2} + c_2 \, (m_s^2 + m_d^2) + c_3 \, m_s \, m_d \right)\, S - \frac{c_{12}}{a}\, (m_s + m_d) \, P - c_4 O_4 \right] \ ,
\ee
where $O_{CM}=16 \pi^2 Q_g^+ = g\, \bar s\, \sigma_{\mu\nu} G_{\mu\nu} d$, $S = \bar s\, d$,  $P = \bar s\, i \,\gamma_5 \, d$ and $O_4 = \Box(\bar s\, d)$.
Note, in particular, that the quadratically divergent mixing of the CMO with the scalar density $S$ is common to any regularization, whereas the mixing with the pseudoscalar density $P$ (softened by the proportionality to the quark masses) is peculiar of twisted mass fermions, and it is a consequence of the non conservation of parity.

The determination of power divergent coefficients, controlling the mixing with operators of lower dimension, cannot rely on perturbation theory~\cite{Maiani:1991az}. The reason is that potential non analytic contributions to these coefficients, like those proportional to powers of $\frac{1}{a} \exp(-1/(\beta_0 g^2)) \sim \Lambda_{QCD}$, do not appear in the perturbative expansion. Therefore, in order to evaluate the coefficients $c_{13}$ and $c_{12}$  in Eq.~(\ref{eq:reno}) we developed a non-perturbative strategy. On the other hand, we computed the other coefficients of Eq.~(\ref{eq:reno}), i.e. the multiplicative renormalization factor $Z_{CM}$ and the coefficients $c_i$ with $i=2,3,4$, in perturbation theory at one-loop. The details of this perturbative calculation, which involves 2- and 3-point functions of the CMO and addresses the whole mixing pattern, including the mixing with the non gauge invariant operators vanishing by the equation of motion, have been presented by Marios Costa in a poster at this conference~\cite{Marios}. We do not discuss it any further here, but present in Table~\ref{tab:coeff} the numerical results for the coefficients obtained at the 3 values of the coupling considered in the numerical simulation, namely $\beta=1.90, \ 1.95, \ 2.10$. 
\begin{table}[t]
\begin{center}
\begin{tabular}{||c||c|c|c|c|c|c|c||}
\hline
$\beta$ & $Z_{CM}$ & $c_2$ & $c_3$ & $c_4$ & $c_{12}$ & $c_{13}$ & $c_{13}$ (non-pert.)  \\
\hline
1.90 & 1.78 & 0.15 & 0 & 0 & 0.085 & 0.96 & 0.8977(2) \\
1.95 & 1.75 & 0.10 & 0 & 0 & 0.083 & 0.94 & 0.8769(4) \\
2.10 & 1.68 & -0.04& 0 & 0 & 0.077 & 0.87 & 0.8165(8) \\
\hline   
\end{tabular}
\end{center}
\normalsize
\caption{\it Values of the multiplicative renormalization factor $Z_{CM}$, in the $\msbar$ scheme at the scale $\mu=$ 2 {\rm GeV}, and of the mixing coefficients $c_i$ of Eq.~(\protect\ref{eq:reno}). The results are obtained at the three values of the lattice coupling $\beta$ from one-loop perturbation theory~\protect\cite{Marios}, except for $c_{13}$ (non-pert.) in the last column which is obtained non-perturbatively from Eq.~(\protect\ref{eq:c13}). The perturbative results have been evaluated using the bare coupling $g_0=6/\beta$ for the power divergent coefficients $c_{12}$ and $c_{13}$ and the boosted coupling $\tilde g^2=g_0^2/U_P$ for the other coefficients, where $U_P$ is the average plaquette.}
\label{tab:coeff}
\end{table}
We also give in the table the one-loop results for the power divergent coefficients $c_{12}$ and $c_{13}$, though their accuracy is expected to be limited, as well as the precise results for $c_{13}$ obtained non-perturbatively from Eq.~(\ref{eq:c13}) below. The non-perturbative determination of $c_{12}$ is in progress, and the results will be presented elsewhere. It can be seen from Table~\ref{tab:coeff} that the coefficients $c_3$ and $c_4$ are found to be zero at one loop, the coefficient $c_2$ which starts at ${\cal O}(g^2)$ is rather small, whereas the multiplicative renormalization factor $Z_{CM}$ receives at one loop a sizable correction ($\sim 70\%$).

Since the the scalar and pseudoscalar densities are proportional to the four-divergence of the vector and axial-vector currents respectively, their appearance in Eq.~(\ref{eq:reno}) does not affect the physical matrix elements of the CMO, in which four-momentum is conserved between initial and final state. In this study, however, we evaluate $B_{CMO}^{K\pi}$ through the unphysical matrix element $\langle \pi (p_\pi)| O_{CM} | K (p_K)\rangle$, with $p_K \ne p_\pi$. The power divergent subtraction is thus required. The proper definition of the subtracted CMO is then provided by the chiral representation (\ref{eq:chiral}) and the corresponding expressions~(\ref{eq:me}) for the CMO matrix elements. They show, in particular, that the matrix element of the CMO between external kaon and pion at rest must vanish in the chiral limit, which provides in turn the following subtraction condition 
\be
\label{eq:c13}
\frac{1}{Z_{CM}} \ \lim_{m_s, ~ m_d \to 0} ~ \langle \pi (0) | \, \widehat O_{CM} \, | K (0) \rangle = 
\lim_{m_s, ~ m_d \to 0} ~ \langle \pi (0) | \, O_{CM} -  \frac{c_{13}}{a^2}\, S \, | K (0) \rangle = 0 \ ,
\ee
from which the coefficient $c_{13}$ can be determined non-perturbatively. Note in particular that the operator $O_4 = \Box(\bar s\, d)$, which contributes to the mixing in Eq.~(\ref{eq:reno}), does not appear on the r.h.s. of Eq.~(\ref{eq:c13}) since its matrix element between $K$ and $\pi$ at rest, being proportional to $(M_K-M_\pi)^2$, vanishes in the chiral limit. 

Once the coefficient $c_{13}$ has been determined from Eq.~(\ref{eq:c13}), the power divergent mixing of the CMO with the pseudoscalar density allowed within the twisted mass regularization can be subtracted non-perturbatively by requiring the vanishing of the parity violating matrix elements of the CMO. Specifically, we imposed
\be
\label{eq:c12}
\frac{1}{Z_{CM}} \ \langle 0 | \, \widehat O_{CM} \, | K \rangle = \langle 0 |\, O_{CM} - \frac{c_{13}}{a^2}\, S - \frac{c_{12}}{a}\, (m_s + m_d) \, P \, | K \rangle = 0
\ee
(up to terms of ${\cal O}(a)$), from which the coefficient $c_{12}$ can be determined.

The non-perturbative results obtained for $c_{13}$ from Eq.~(\ref{eq:c13}) are collected in the last column of Table~\ref{tab:coeff}. Details of the analysis are given in~\cite{Marios}. These non-perturbative results differ by about 7\% with respect to the one-loop predictions, see Table~\ref{tab:coeff}, and the bulk of the difference is compatible with being a correction of ${\cal O}(g^4)$. Thus, genuine non-perturbative contributions to $c_{13}$ are likely to be small, even though a firmer conclusion in this sense would require the calculation of $c_{13}$ at two loops at least. 

As far as the coefficient $c_{12}$ is concerned, its non-perturbative determination based on Eq.~(\ref{eq:c12}) is still in progress. Nevertheless, preliminary results indicate that its size is correctly predicted by the one-loop estimate presented in Table~\ref{tab:coeff}. This estimate is smaller by one order of magnitude than $c_{13}$. In addition, the corresponding operator is proportional to one power of the quark masses, with $a (m_s + m_d)\sim 0.02$ in our simulation. For this reason, the subtraction of the linear divergence in Eq.~(\ref{eq:reno}) turns out to be numerically negligible and, for the time being, we performed it by using for $c_{12}$ its one-loop perturbative determination.

\begin{wrapfigure}{r}{0pt}
\includegraphics[scale=0.3]{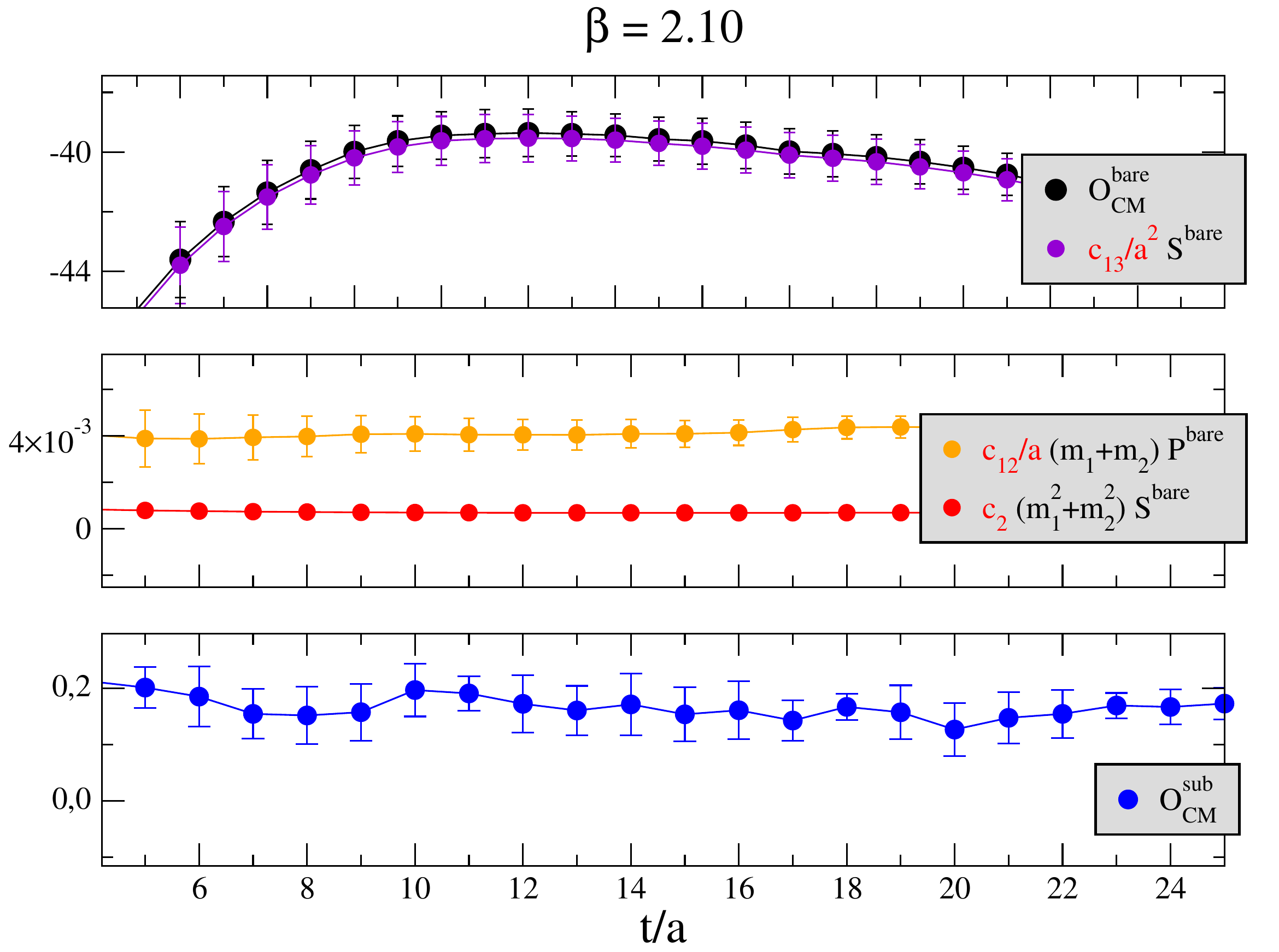}
\vspace{-0.2cm}\caption{\it Time plateaus of the $K\to\pi$ matrix elements of the bare operators (top and center) which contribute to the matrix elements of $Z_{CM}^{-1} \, \widehat O_{CM}$ (bottom). The results refer to $\beta=2.10$ and to a light quark mass corresponding to $M_\pi \simeq 210$ MeV.}
\label{fig:cmo0}
\end{wrapfigure}
The size of the various $K\to\pi$ matrix ele\-ments of bare operators contributing, according to Eq.~(\ref{eq:reno}), to the matrix element of the renormalized CMO can be inferred from Fig.~\ref{fig:cmo0}, which shows the time plateaus from which these matrix elements are extracted. The results refer to $\beta=2.10$. As can be seen from the plots, the matrix elements of the bare CMO and of the leading power divergence, $(c_{13}/a^2)\, S$, are of the same size. 
The subtraction is at the level of 99.5\%. The other two contributions, i.e. the matrix elements of the operators $(c_{12}/a)\, (m_s+m_d)\,P$ and $c_2 \, (m_s^2 + m_d^2)\, S$, are smaller by four orders of magnitude. The operators $m_s\,m_d\, S$ and $O_4$ do not contribute at one loop, by having vanishing coefficients. In the lower panel of Fig.~\ref{fig:cmo0} we show the resulting matrix element of the subtracted operator, $Z_{CM}^{-1} \, \widehat O_{CM}$. Despite being the outcome of a large numerical subtraction, the result is clearly different from zero and the matrix element turns out to be determined quite precisely.

\section{Matrix elements of the chromomagnetic operator}
\begin{figure}[b]
\centering
\includegraphics[scale=0.28]{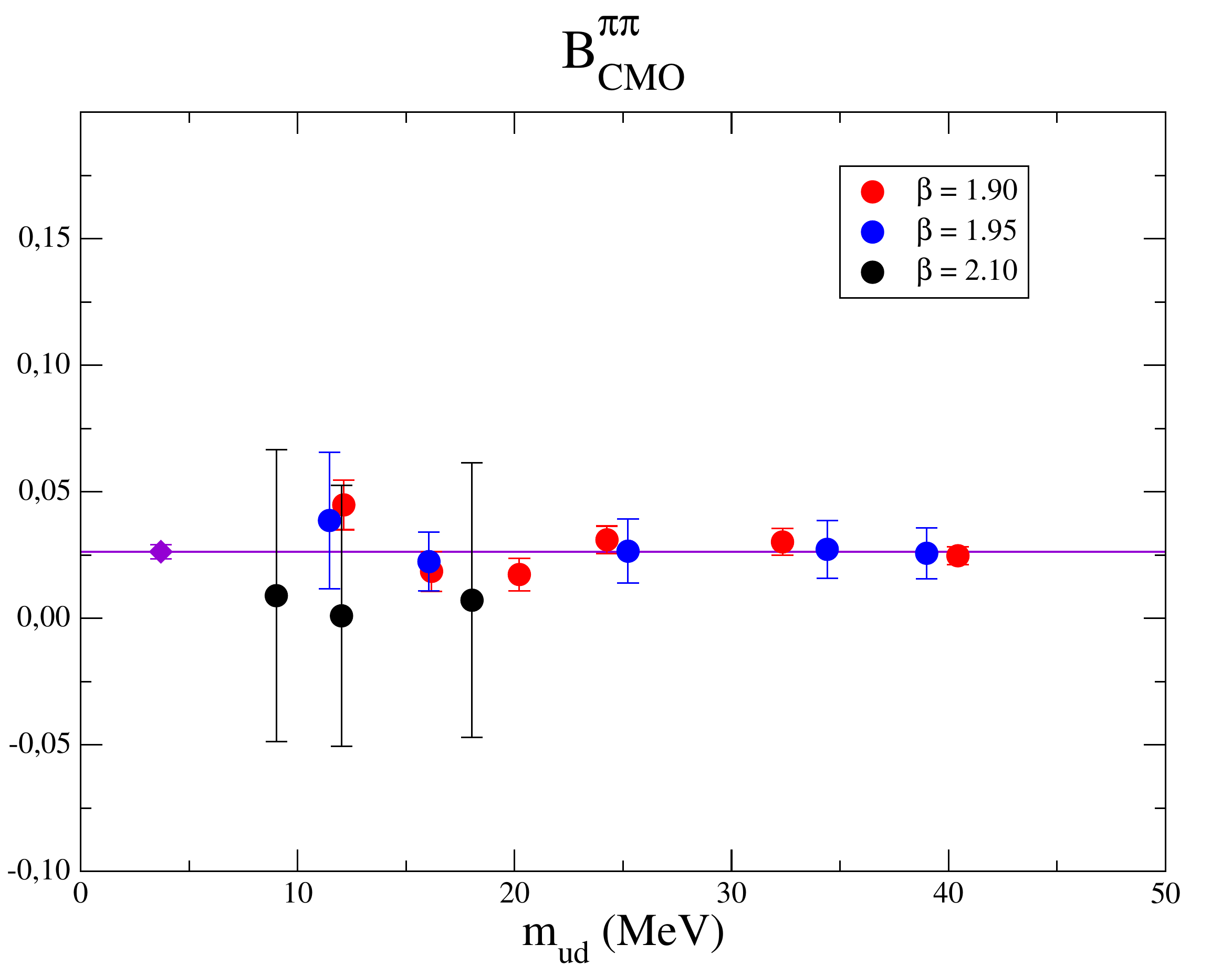}
\includegraphics[scale=0.28]{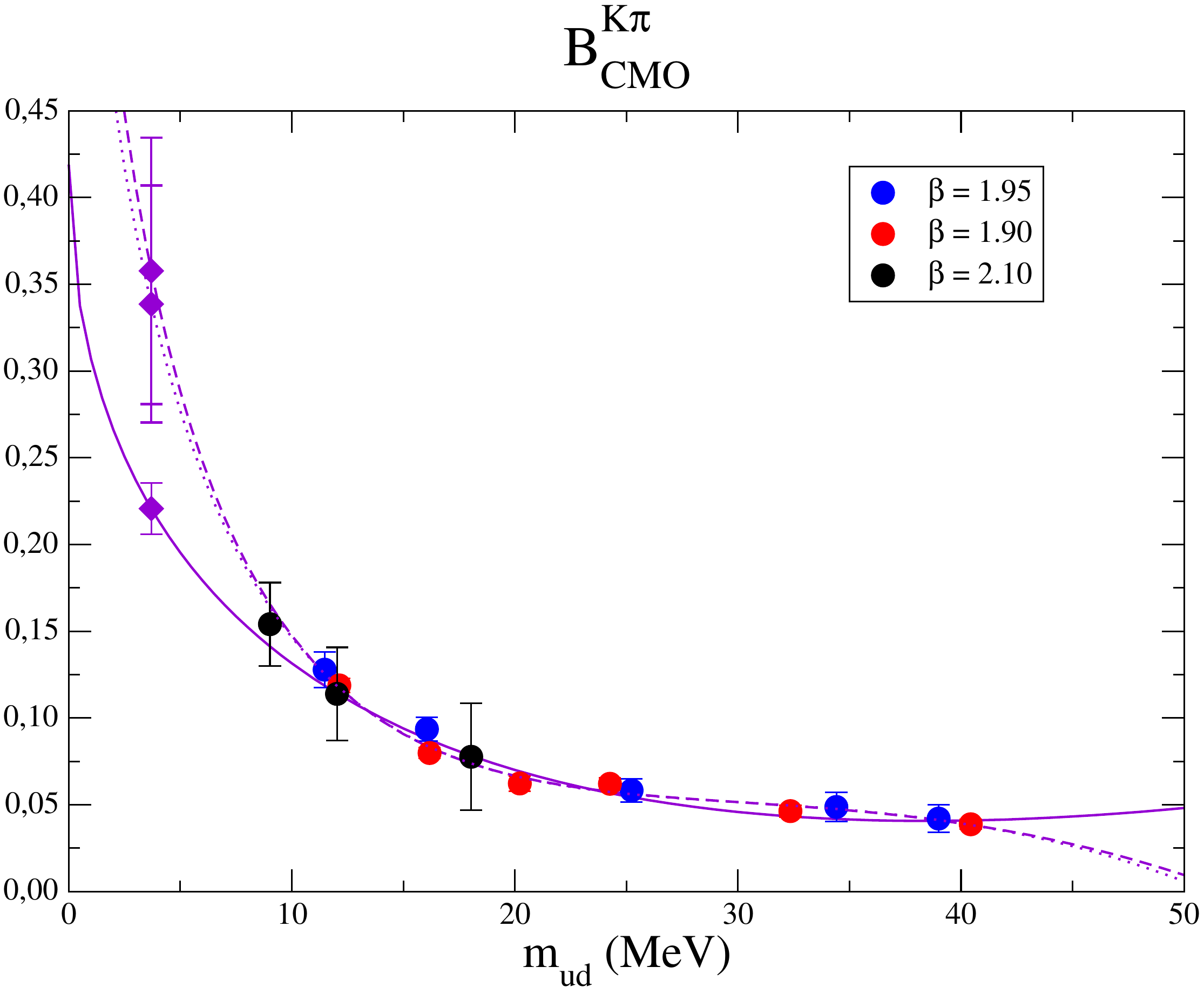}
\caption{\it Values of the B-parameters $B_{CMO}^{\pi\pi}$ (left) and $B_{CMO}^{K\pi}$ (right) as a function of the light quark mass. On the right plot, solid, dashed and dotted lines represent the fits according to Eq.~(\protect\ref{eq:bcmofit}) with $\Delta(m_{ud})=0$, $\Delta(m_{ud})\sim m_{ud}^{3/2}$ and $\Delta(m_{ud})\sim m_{ud}^{2}$ respectively. The diamonds represent the results of the extrapolation to the physical up-down average quark mass.}
\label{fig:Bextra}
\end{figure}

By having defined the properly renormalized CMO, we are now in the position of computing its matrix elements extrapolated at the physical quark masses. In the right plot of Fig.~\ref{fig:Bextra} we show the results for the B-parameter $B_{CMO}^{K\pi}$, as a function of the light quark mass $m_{ud}$, extracted from the $K\to\pi$ matrix element with external mesons at rest, according to Eq.~(\ref{eq:me}). 
For comparison, we also show in the left plot the parameter $B_{CMO}^{\pi\pi}$, which is obtained from $B_{CMO}^{K\pi}$ by considering the limit of degenerate valence quarks, i.e. $m_s = m_d = m_{ud}$. Note that, in both cases, the results obtained at the three values of $\beta$ lie on the same curve, showing completely negligible discretization effects within the errors. This is most likely a beneficial effect of the subtraction~(\ref{eq:c13}). Moreover, in the $\pi\to\pi$ case, the B-parameter is also found to be practically independent of the light quark mass, indicating that the chiral behavior predicted by ChPT at LO is well verified for this matrix element.

Chiral corrections beyond the LO are clearly visible, instead, in the $K\to\pi$ matrix element and the B-parameter $B_{CMO}^{K\pi}$ shown in the right panel of Fig.~\ref{fig:Bextra} exhibits a large dependence on the light quark mass. In order to extrapolate to the physical mass, we have then taken into account NLO as well as higher order chiral corrections and fitted the B-parameter according to
\bea
B_{CMO}^{K\pi}(m_{ud}) &=& \left(B_{CMO}^{K\pi}\right)^{LO} \left[ 1 + \alpha \, M_K^2  + \beta \, M_\pi^2  + \gamma \,  p_K\cdot p_\pi + \Delta \right]\vert_{\vec p_K =\vec p_\pi = 0} = \nonumber \\
&=& \left(B_{CMO}^{K\pi}\right)^{LO} \left[ \alpha^\prime + \beta^\prime \, m_{ud}  + \gamma^{\,\prime} \,  m_{ud}^{1/2} + \Delta(m_{ud}) \right] \ ,
\label{eq:bcmofit}
\eea
where, in the second line, we have made explicit the dependence on the light quark mass. The function $\Delta$ includes chiral corrections beyond the NLO. We have performed fits with $\Delta=0$, $\Delta \sim m_{ud}^{3/2}$ and $\Delta \sim m_{ud}^{2}$ obtaining in all cases a good description of the lattice data (see Fig.~\ref{fig:Bextra}). The extrapolation to the physical point, in the three cases, leads to 
\be
B_{CMO}^{K\pi}=0.22(2) \, \vert_{\Delta=0} \quad , \quad 
B_{CMO}^{K\pi}=0.36(8) \, \vert_{\Delta \sim m_{ud}^{3/2}}  \quad , \quad  
B_{CMO}^{K\pi}=0.34(7) \, \vert_{\Delta \sim m_{ud}^{2}} \ .
\label{eq:chfits}
\ee
As preliminary result for the B-parameter of the CMO we finally quote the value
\be
B_{CMO}^{K\pi} \ = \ 0.29\,(9)_{stat+fit}\,(6)_{PT} \ = \ 0.29\,(11) \ , 
\label{eq:final}
\ee
where the first error includes the statistical error and the uncertainty due to the chiral extrapolation, estimated from the spread of the results in Eq.~(\ref{eq:chfits}), while the second error accounts for the perturbative uncertainty in the one-loop determination of the multiplicative renormalization factor $Z_{CM}$, see Table~\ref{tab:coeff}. Our result~(\ref{eq:final}) represents the first lattice QCD determination of a matrix element of the CMO. The comparison with the estimate $B_{CMO}^{K\pi}\sim 1-4$ currently adopted in phenomenological analyses~\cite{Buras:1999da} indicates that the uncertainty on this quantity has been significantly reduced.

\end{document}